\documentclass{epl}
\usepackage{epsfig}

\title{New mechanism for impurity-induced step bunching}
\author{Joachim Krug}
\institute{Fachbereich Physik, Universit\"at Essen,
    D-45117 Essen, Germany
}

\begin{document}

\maketitle

\begin{abstract}
Codeposition of impurities during the growth of a vicinal surface
leads to an impurity concentration gradient on the terraces, which
induces corresponding gradients in the
mobility and the chemical potential of the adatoms.
Here it is shown that the two types of gradients
have opposing effects on the stability of the surface: Step 
bunching can be caused by impurities which either lower the adatom
mobility, or increase the adatom
chemical potential. In particular, impurities acting as 
random barriers (without affecting the adatom binding)
cause step bunching, while for  
impurities acting as random traps the combination of the two
effects reduces to a modification of the attachment
boundary conditions at the steps. In this case attachment to descending steps,
and thus step bunching, is favored if the impurities bind 
adatoms more weakly than the substrate.

\end{abstract}

\emph{Introduction.}
Step bunching is a morphological instability of a vicinal crystal
surface, in which a regular train of equally spaced steps
separates into regions of high step density -- the step bunches --
and large flat terraces. The process can be driven energetically
by an attractive step-step interaction, or by a variety of kinetic
mechanisms, which all share the common feature of breaking the 
symmetry between the ascending (upper) and descending (lower)
step bordering the vicinal terrace \cite{PimpVill}. 
In growth or sublimation,
the symmetry breaking is provided by the different kinetic rates
for the attachment and detachment of adatoms \cite{Schwoebel} 
(or some other species required for growth \cite{Pimp}) 
at the upper and the lower
step; step bunching occurs under growth 
if atoms attach preferentially to the descending step. 
For electromigration-induced step bunching, the asymmetry
is introduced by the electric field, and the step train
is unstable if the adatom motion is biased in the down-step
direction \cite{Stoyanov}.

It has been appreciated for a long time that in many cases
step bunching must be attributed to the presence of impurities
\cite{Cabrera}. 
The traditional view is that impurities pin the steps
\cite{Kandel}. Once a step
is slowed down relative to its neighbors, more impurities accumulate
in front of it and delay it even further, leading to a feedback
mechanism which drives the instability \cite{HMK}. A different
kind of impurity-mediated step bunching was suggested in recent
work on $\mathrm{Si}_{1-y} \mathrm{C}_y$ layers grown on Si(100)
by molecular beam epitaxy, in which C plays the role of a 
codeposited impurity \cite{Croke}. The key observation is that
different parts of the vicinal terrace have been exposed to the impurity
flux for different durations. Therefore the impurity concentration 
is smallest on the freshly created part near the descending step, and
largest near the ascending step. To the extent that the impurities
couple to the energetics and kinetics of the adatoms on the terrace,
this causes corresponding gradients in the adatom chemical potential and
mobility which break the symmetry between ascending and descending
steps, and hence may lead to step bunching\footnote{A similar mechanism
for step \emph{equalization} was proposed in Ref.\cite{Pond}.}. 

For the SiC system,
the experimentally observed step bunching could be reproduced 
in simulations in which the Si-C binding was assumed to be weaker
than the Si-Si binding. This was interpreted in terms of an 
increase of the adatom mobility due to the impurities: The low
concentration of impurities near the descending step was argued
to lead to an accumulation of adatoms in these low-mobility regions,
and hence to a preferential attachment to the descending step. 
However, the adatom flux onto a step depends not only on the 
adatom density gradient, but also on the adatom mobility, which is lower
near the descending step. The explicit calculations presented below
show that the latter effect overcompensates the increase in the
adatom concentration gradient. Impurities which increase the adatom mobility
are found to stabilize the step train, while step bunching 
is induced if the adatoms are slowed down by the adsorbates.

\begin{figure}
\centerline{\epsfig{file=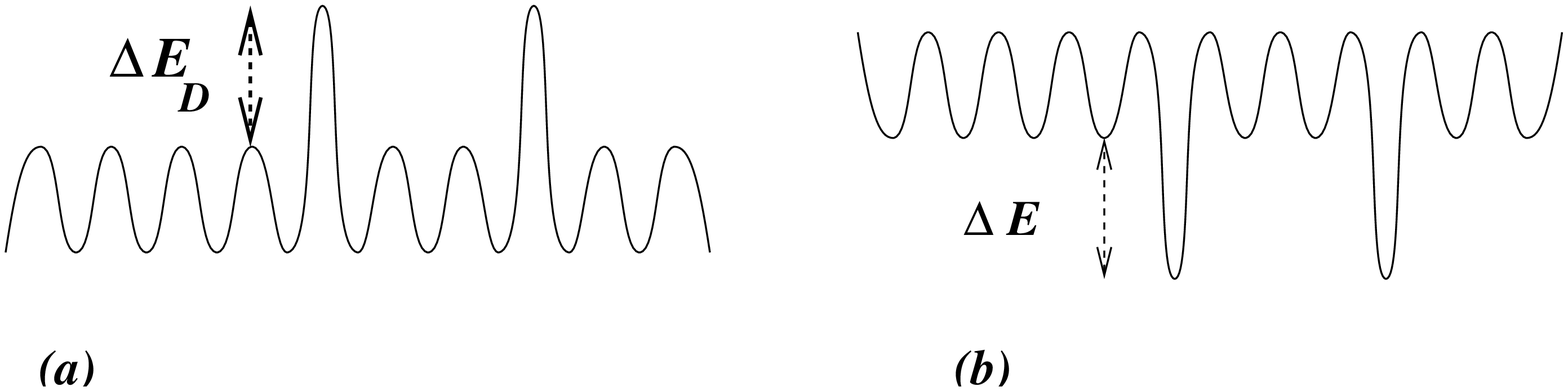,width=12.cm}} 
  \caption{Adatom energy landscapes used in this work. (a) Random barriers:
Impurities modify the adatom diffusion barrier by $\Delta E_D$ while leaving
binding energies unchanged. 
(b) Random traps: The adatom binding energy and the 
diffusion barrier for escape from an impurity site is modified by  
$\Delta E$, while the transition state energies remain unchanged.}
\label{potentials}
\end{figure}

On the other hand, the chemical potential gradient induced by the impurities
acts in the opposite direction. For impurities which bind the adatoms
more strongly than the clean substrate, and which would therefore be
expected to lower the adatom mobility, the chemical potential is decreased
near the ascending step edge, where the impurities accumulate. This 
implies an uphill force on the adatoms, which, as is well known from studies
of electromigration-induced step bunching, stabilizes the step train
\cite{Stoyanov}. Similarly, for impurities that bind more weakly than
the substrate, the chemical potential gradient is destabilizing and
the mobility gradient stabilizing. The net outcome of the two competing
effects can be determined only if the modification of the adatom potential
energy landscape caused by the impurities is precisely specified.
Two limiting cases will be considered in detail: Random barriers which
modify only the adatom mobility, and random traps for which
binding energies and diffusion barriers are modified by the same
amount (see Fig.\ref{potentials}). Random barriers and random traps are
standard models in the theory of diffusion in disordered media
\cite{Haus}.        

\begin{figure}
 \centerline{\epsfig{file=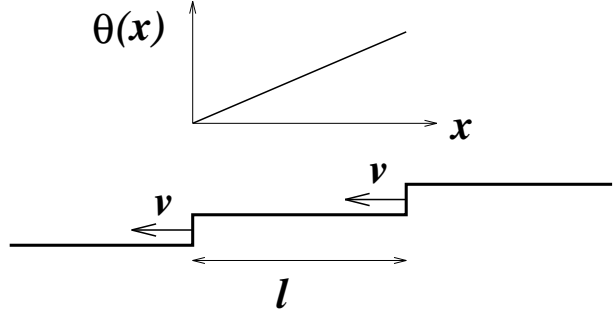,width=8.cm}} 
  \caption{Schematic of the unperturbed step train 
and the linear impurity concentration
profile (\ref{thetax}).}
  \label{steptrain}
\end{figure}

\emph{Model and general solution.}
Figure \ref{steptrain} illustrates the geometry employed in the calculation.
I consider a train of straight steps with spacing $l$. The deposition flux
is $F$ and the impurity flux $F'$. Impurities are immobile, they
do not desorb, and they are incorporated into the crystal
when a step moves over them. The steps move with speed $v = F l$.
The exposure time at a distance $x$ from the descending step is 
$x/v$, hence the stationary impurity coverage profile is 
\begin{equation}
\label{thetax}
\theta(x) = F' x/v = \phi x/l,
\end{equation} 
where $\phi = F'/F$ is the flux ratio.
The spatial variation of the impurity concentration implies
a corresponding variation of the effective chemical potential
$\mu_{\mathrm{eff}}(x)$ and the adatom diffusion coefficient
$D(x)$, which will be specified later. Assuming that the
adatom concentration $n(x)$ adapts rapidly to changes in the step
spacing and in the impurity profile, it can be computed from the
inhomogeneous, stationary diffusion equation \cite{Penev} 
\begin{equation}
\label{diffusion}
\frac{d}{dx} D(x) \frac{d}{dx} \left[ 
\frac{dn}{dx} + \beta n \frac{d \mu_{\mathrm{eff}}}{dx} 
\right] + F = 0,
\end{equation}
where $\beta = 1/k_{\mathrm{B}} T$. This is supplemented by boundary
conditions for the mass fluxes $j_-$ and $j_+$ to the
descending ($x=0$) and ascending ($x=l$) step \cite{Schwoebel,Stoyanov},
\begin{equation}
\label{bc1}
j_- = D(0)[n'(0) + \beta n(0) \mu_{\mathrm{eff}}'(0)] = k_- n(0)
\end{equation}
\begin{equation}
\label{bc2}
j_+ = - D(l) [n'(l) + \beta n(l) \mu_{\mathrm{eff}}'(l)] = k_+ n(l).
\end{equation} 
The attachment rates $k_-$, $k_+$ are chosen such
that the attachment probability is symmetric in the absence of impurities.
Specifically, I will consider two types of boundary conditions:
Type I with $k_- = k_+ = k$ (attachment rates independent of the impurity
concentration) and type II with $k_-/D(0) = k_+/D(l) = \lambda^{-1}$
(attachment rate proportional to the adatom mobility at the step). 

The mass fluxes $j_-$ and $j_+$ govern the 
dynamics of the vicinal surface.
The two are related through mass conservation, 
$j_- + j_+ = F l$. To probe the stability of the uniform step train,
consider a period-2
perturbation in which the length of every second terrace
is increased by an amount $\epsilon$ and every second terrace length is
decreased by $\epsilon$. In the absence of impurities 
this does not affect the speed of the steps, since the attachment rates
$k_\pm$ are symmetric, and hence
the total flux feeding each step remains
$Fl$. The impurity profile associated with the perturbed step train
is therefore still given by (\ref{thetax}). When the coupling of the
impurities to the adatom concentration is turned on, the larger 
terraces either shrink, restoring the uniform step train, or grow,
leading to step doubling and, eventually, to step bunching.
The large terraces shrink, if the speed of the corresponding
ascending step is larger than that of the descending step, i.e. if
\begin{equation}
\label{shrink}
j_+(l+\epsilon) + j_-(l-\epsilon) > j_+(l-\epsilon) + j_-(l + \epsilon).
\end{equation} 
In the limit $\epsilon \to 0$ this becomes
$d j_+/dl > d j_-/l$ or, using mass conservation, 
\begin{equation}
\label{stability}
d j_-(l) /dl < F/2,
\end{equation}     
which has to be evaluated at fixed $\theta(x)$, i.e. without 
taking into account the $l$-dependence of the impurity 
concentration gradient.
When the stability criterion (\ref{stability}) is violated, the growth
rate of the perturbation determines the time scale for step bunching, which
is given by 
\begin{equation}
\label{tau}
\tau = (4 dj_-/dl - 2F)^{-1}.
\end{equation}

Following \cite{Penev}, the mass flux $j_-$ to the descending step is  
obtained from Eqs.(\ref{diffusion},\ref{bc1},\ref{bc2}) in the general form
\begin{equation}
\label{jgeneral}
j_- = Fl \left( \frac{M_1 + 1/\tilde k_+}{M_0 + 1/\tilde k_+ + 
1/\tilde k_-} \right),
\end{equation}
where 
\begin{equation}
\label{M}
M_\nu = l^{-\nu} \int_0^l dx \; x^\nu e^{\beta \mu_{\mathrm{eff}}(x)} 
D(x)^{-1}
\end{equation}
and 
\begin{equation}
\label{ktilde}
\tilde k_- = e^{-\beta \mu_{\mathrm{eff}}(0)} k_-, \;\;\;
\tilde k_+ = e^{-\beta \mu_{\mathrm{eff}}(l)} k_+. 
\end{equation}

\emph{Random barriers.}
Consider first the case where the impurities affect \emph{only} the
mobility of the adatoms. This corresponds to the random barrier energy
landscape illustrated in Fig.\ref{potentials} (a), where the binding
energies remain unaffected while the diffusion barriers are modified
by an amount $\Delta E_D$, which can be positive 
(as in Fig.\ref{potentials} (a)) or negative. An exact analytic expression
for the effective diffusion coefficient in two dimensions is not
available for the random barrier model \cite{Haus}, but some general
conclusions can be drawn from Eqs.(\ref{jgeneral},\ref{M}), which are
to be evaluated with $\mu_{\mathrm{eff}} = 0$. Since $D(x)$ is a 
monotonic function of $x$, we have that $l/D(0) \leq M_0 \leq 
l/D(l)$, $l/2 D(0) \leq M_1 \leq l/2 D(l)$ and $M_1/M_0 \geq 1/2$
for $\Delta E_D > 0$, and the converse inequalities for $\Delta E_D < 0$.
Using these relations it is straightforward to prove that attachment
is primarily to the descending step ($j_- \geq Fl/2 \geq j_+$) 
when $\Delta E_D > 0$, and to the ascending step when $\Delta E_D < 0$, 
for both types of boundary conditions.      

For the explicit calculation of $j_-$ I use the expression
\cite{Frankl,Kotrla00} 
\begin{equation}
\label{Dx}
D(x) = \frac{D_0}{1 + (e^{\beta \Delta E_D} - 1) \theta(x)}
\equiv \frac{D_0}{1 + b x},
\end{equation}
which is exact for the \emph{one-dimensional} 
random barrier model \cite{Haus}. Here
$D_0$ is the diffusion coefficient on the clean surface, and
$b = \phi (e^{\beta \Delta E_D} - 1)/l$
is a dimensionless parameter describing the strength and the
sign of the mobility gradient\footnote{Note that $b l > - \phi$ with
$\phi = F'/F \ll 1$, hence $D(x)$ is positive and finite everywhere.}.  
For type I boundary conditions the flux to the descending step reads
\begin{equation}
\label{kinetic}
j_- = \frac{Fl}{2} \; \frac{1 + l/2 \lambda_0 + b l^2/3 \lambda_0}{1 + 
l/2 \lambda_0 + b l^2 /4 \lambda_0},
\end{equation}
where $\lambda_0 = D_0/k$. Taking the derivative
of (\ref{kinetic}) at fixed $b$, one finds that the stability
criterion (\ref{stability}) is satisfied (violated) when $b < 0$
($b > 0$). Thus step bunching is induced by impurities which slow
down adatom diffusion ($\Delta E_D > 0$). 
The instability is directly
linked to the preferential feeding of the steps from above, i.e.
the stability criterion (\ref{stability}) is equivalent to
$j_- < Fl/2$. This need not be true in general.

\begin{figure}
 \centerline{\epsfig{file=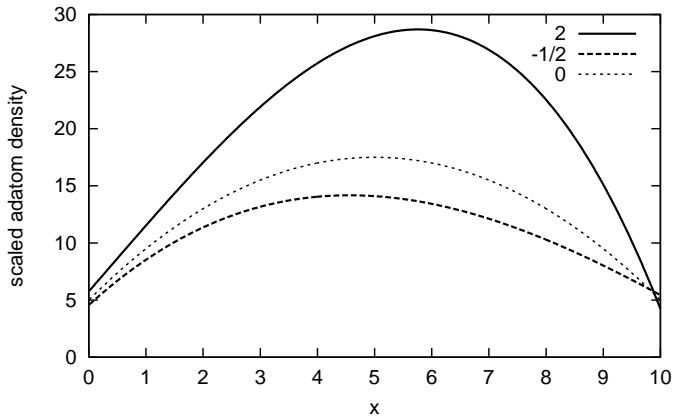,width=9.cm}}
  \caption{Adatom density profile (\ref{density}) 
for $l/\lambda_0 = 10$
and $bl = $ 2, --1/2 and 0, respectively. The adatom density has
been scaled by the overall factor $F/D_0$.}
  \label{densfig}
\end{figure}

The corresponding adatom density profile is given by 
\begin{equation}
\label{density}
n(x) = \frac{F}{D_0} \left(A (\lambda_0 + x) + \frac{1}{2}(A b - 1) x^2 -
\frac{1}{3} b x^3 \right),
\end{equation}
where $A = j_-/F = k n(0)/F$. The examples depicted in Fig.\ref{densfig}
show how the density maximum shifts towards the ascending (descending)
step for $b > 0$  ($b < 0$), as would be expected 
intuitively. Consequently the density gradient is enhanced near
the ascending (descending) step. As was mentioned already,
this effect is however overcompensated by the spatial dependence
of the adatom mobility. The boundary values of the adatom
density vary in the opposite direction to the density gradients, so that 
the mass flux is predominantly
to the descending (ascending) step for $b > 0$ ($b < 0$). 

For type II boundary conditions the flux to the descending step is given by
\begin{equation}
\label{kinetic2} 
j_- = \frac{Fl}{2} \; \frac{1 + l/2 \lambda_0 + b l + b l^2/3 \lambda_0}{1 + 
l/2 \lambda_0 + b l/2 + b l^2 /4 \lambda_0},
\end{equation} 
which behaves similar to (\ref{kinetic}):
Attachment is primarily to the descending (ascending) step and the step
train is unstable (stable) when $b > 0$ ($b < 0$). This shows that
step bunching caused by random barrier impurities is 
a robust phenomenon which is independent of the detailed model assumptions.

\emph{Random traps.}
In general, the influence of the impurity-induced chemical
potential gradient has to be taken into account as well.
The effective chemical potential $\mu_{\mathrm{eff}}$ is obtained
from a thermodynamic argument. We assume that the impurities modify the
adatom binding energy by an amount $\Delta E_b$, $\Delta E_b > 0$
corresponding to stronger binding. The equilibrium adatom density
in a region with impurity concentration $\theta$ is then given
by $n_0(\theta) = n_0(0) (1 - \theta + \theta e^{\beta \Delta E_b})$,
since the occupation of impurity sites is enhanced or suppressed
by the Boltzmann factor $e^{\beta \Delta E_b}$. Writing 
$n_0(\theta) = n_0(0) e^{-\beta \mu_{\mathrm{eff}}}$ and inserting
the linear impurity profile (\ref{thetax}), we obtain
\begin{equation}  
\label{mueff}
\mu_{\mathrm{eff}}(x) = - k_B T \ln(1 + f x)
\end{equation}
where $f = (e^{\beta \Delta E_b} - 1)(\phi/l)$ is the analogue of 
$b$ in (\ref{Dx}). The effective force entering (\ref{diffusion})
is then given by $- \beta d\mu_{\mathrm{eff}}/dx = f/(1 + fx)$,
which points uphill (stabilizing the step train \cite{Stoyanov}) when
$\Delta E_b > 0$, and downhill (destabilizing the step train) when 
$\Delta E_b < 0$.

In contrast to the impurity-induced mobility gradient, the
gradient in the chemical potential cannot occur in isolation
(at constant $D(x)$), because an energy landscape in which
the impurities modify the adatom binding energy without affecting
the diffusion barriers is not conceivable. 
A simple yet realistic situation where both effects are present
simultaneously is provided by the random trap model, illustrated
in Fig.\ref{potentials}(b).
In this model
it is assumed that the transition states between diffusion sites remain
unaffected by the impurities, so that the binding energies
and the diffusion barriers (for jumps away from an impurity
site) are modified by the same amount, $\Delta E_b = \Delta E_D 
\equiv \Delta E$. The effective diffusion coefficient is then
given \emph{exactly} by (\ref{Dx}) in all dimensions \cite{Haus}. Combining 
(\ref{mueff}) and (\ref{Dx}) with $b = f$, the integrand 
$e^{\beta \mu_{\mathrm{eff}}} D(x)^{-1}$ in 
(\ref{M}) is seen to become constant. Hence $M_0 = l/D_0$, 
$M_1 = l/2 D_0$ and (\ref{jgeneral}) reduces to the familiar
expression \cite{Schwoebel} for the clean surface, but 
with modified attachment
rates $\tilde k_-=k_-$, $\tilde k_+ = k_+(1+bl)$. 

This is a consequence of the fact
that for \emph{unbiased} potential landscapes, in which the jump
rates away from a given site are everywhere symmetric, the 
inhomogeneous diffusion equation (\ref{diffusion}) can be written
as \cite{Landauer,Collins}
\begin{equation}
\label{diffusion2}
\frac{d^2}{dx^2} [D(x) n(x)] + F = 0,
\end{equation}
which implies that $\tilde n(x) \equiv D(x) n(x)$ satisfies
a diffusion equation with constant coefficients
and standard boundary conditions $D(0) \tilde n'(0) = k_- \tilde n(0)$,
$D(l) \tilde n'(l) = - k_+ \tilde n(l)$.
For type II boundary conditions $k_+/D(l) = k_-/D(0)$ so that 
attachment remains symmetric in the presence of impurities, and
the two competing impurity effects precisely cancel. In contrast,
for type I boundary conditions the impurities are seen to induce
a preference for attachment at the ascending step for $\Delta E > 0$,
and at the descending step for $\Delta E < 0$. This implies a tendency
towards step bunching for $\Delta E < 0$. The effect is however quite
feeble, since $\vert b l \vert < \phi \ll 1$ for $\Delta E < 0$.  

\emph{Summary.}
In conclusion, I have described a novel mechanism through which codeposited
adsorbates may destabilize a growing vicinal surface. The kinetic
and energetic couplings between adsorbate atoms and
adatoms were shown to have competing effects. 
Impurities which slow down the adatom diffusion without affecting
adatom binding energies (random barriers) generically cause step bunching.
When the impurities act as random traps, and provided 
the attachment rates at the steps are not modified by the impurities
(type I boundary conditions),
the net result of the two effects was found to be destabilizing
for impurities that bind adatoms more weakly than the substrate. 
This is consistent with the simulations of SiC growth \cite{Croke}
that inspired the present study. However, the precise cancellation
of the two effects for type II boundary conditions also suggests
that predicting the stability of the surface may be difficult
if the details of the adsorbate-adatom-interaction are not known.

The observation that barrier-like and trap-like impurities may have
qualitatively different effects on the stability of a growing
surface, because they affect the symmetry of the surface diffusion process
in different ways, was made previously in the context of growth on
singular surfaces \cite{Amaral}. It is interesting to note that also in
this case the barrier-like impurities induce a downhill diffusion
bias, favoring attachment to the descending step, through a mechanism that
however does not involve any impurity concentration gradient.     

Future work should address the competition between the impurity-induced
instability and the stabilizing effect of conventional step edge
barriers \cite{Schwoebel}. Going beyond the linear stability analysis
presented here will be difficult because the dynamics becomes nonlocal
in time when the impurity profile is nonstationary \cite{HMK}. 
Further underpinning for the proposed
mechanism from KMC simulations would therefore 
be highly desirable.    

\acknowledgments

I am grateful to F. Grosse for pointing out
\cite{Croke} and for highly useful discussions. 
Remarks by D. Kandel and P. Kratzer were also much 
appreciated. This work was supported
by DFG within SFB 237, and by Volkswagenstiftung.

\end{document}